\DeclareFontShape{OML}{cmm}{m}{b}{%
   <-> cmmib10}{}
\DeclareMathAlphabet{\mathbf}{OML}{cmm}{m}{b}
\DeclareSymbolFont{boldletters}{OML}{cmm}{m}{b}
\DeclareMathSymbol{\btau}{\mathord}{boldletters}{28}

\def\pf{p_{\rm f}}
\def\pj{p_{\rm j}}
\def\diag{{\rm diag}}
\def\L{{\mathcal L}}
\def\R{{\mathcal R}}
\def\M{{\mathcal M}}
\def\S{{\mathcal S}}
\documentstyle[12pt,amssymb,array,cite]{article}
\begin{document}
\begin{titlepage}
\begin{center}
{\Large \bf Exact stationary state for a deterministic high speed traffic
model with open boundaries.}
\end{center}
\vskip2cm
\begin{center}
{\large Jan~de~Gier\footnote{\tt degier@maths.anu.edu.au}} \\
{\it  Department of Mathematics}\\
{\it  School of Mathematical Sciences}\\
{\it  Australian National University}\\
{\it  Canberra ACT 0200, Australia}
\end{center}
\vskip1cm
\centerline{\today}
\vskip2cm
\centerline{\bf Abstract}
\vskip5mm
An exact solution for a high speed deterministic traffic model with open
boundaries and synchronous update rule is presented. Because of the strong
correlations in the model, the qualitative structure of the stationary
state can be described for general values of the maximum speed. It is
shown in the case of $v_{\rm max}=2$ that a detailed analysis of this
structure leads to an exact solution. Explicit expressions for the
stationary state probabilities are given in terms of products of
$24\times 24$ matrices. From this solution an exact expression for the
correlation length is derived. 

\end{titlepage}

\section{Introduction}
\label{se:intro}
One dimensional driven diffusive processes have proven to be an
interesting playground for the study of non-equilibrium behaviour
\cite{L85, L99, SZ95, SW98, CSS00}. Of great interest is the fact that the 
study of many stationary state properties has come within reach of
exact analytical methods since the solution of the asymmetric simple
exclusion process (ASEP) with random sequential update and open
boundaries \cite{DDM92, DE93a, SD93a, DEHP93}. An important
analytical tool in the study of these diffusive systems is the matrix
product method, that appeared earlier in the study of lattice animals
\cite{HN83} and the ground states of antiferromagnets \cite{AKLT87,
KSZ91}. Its use in \cite{DEHP93} for the ASEP has boosted a lot of
research on a variety of diffusion models, among which are for example
the ASEP with other updates \cite{H96, RSS96, HP97, RSSS98, ERS99,
dGN99, BPV00}, multi-species models \cite{DJLS93, EFGM95, M96, HS97,
EKKM98,AHR98, ADR98, K99, FJ99, RSS00}, multi-lane traffic
\cite{LPK97} and the partially asymmetric exclusion process \cite{S94,
ER96, BECE99, Sa99}. For recent reviews of many of the exact results
for the ASEP see \cite{DE97, D98, S00}.  

It has been shown for different dynamical update rules, that the
stationary state of a stochastic model can always be written as a
matrix product \cite{KS97,RS97, KS99}, although no proof is given for
the synchronous update. This mere fact by no means solves the problem
of finding the stationary state, but it provides a basis for a
systematic study via the representation theory of non-linear algebras
\cite{ER96, ADR98, AHR98}. In almost all cases studied so far, the
algebra has been quadratic, which is peculiar to systems with nearest
neighbour interactions only. Only for the synchronous ASEP, non
trivial representations of an algebra of higher degree have been used,
i.e. either quartic \cite{ERS99}, where the matrices depend on one
site, or cubic \cite{dGN99} in the case where the matrices depend on two
sites. 

Recently the asymmetric exclusion process with next nearest
neighbour interactions has been studied by various methods
\cite{AS00}, but in the case of open boundaries exact results have
been obtained only on a special line. In an initial attempt to find
exact stationary states for models with long range interactions and
open boundaries, a deterministic high speed asymmetric exclusion model
is studied. Particles are allowed to hop over more than one lattice
spacing per time step, and they enter the system at the left and leave
at the right. Furthermore, the system will be subject to a synchronous
dynamical update rule. For such dynamics the correlations are the
strongest, which is not only interesting from a physical point of
view, but actually helps solving the problem. The correlations are so
strong that one can describe the stationary state qualitatively in a
simple way by identifying so called Garden of Eden states \cite{TE98,
SS98}.    

The exact stationary state is given in matrix product form for the
case where particles may hop over two lattice sites. The matrices
depend on three sites and it is shown that they should satisfy an
algebra of at least quartic degree. The model is solved by making an
Ansatz for the form of the matrices based on the qualitative
observations for the stationary state. It is then shown that the
submatrices in this Ansatz must satisfy certain relations for the
matrix product state to solve the stationary master equation. A
solution of these relations is found with the help of explicit
calculations for small systems.  

This paper is organised as follows. The model is defined in detail in
Section \ref{se:model}. The exact stationary state is calculated in
Section \ref{se:state} and some results on the phase diagram are
discussed in Section \ref{se:results}.

\section{Definition of the model}
\label{se:model}
In this paper we study a one dimensional asymmetric particle hopping
model where particles in the bulk hop to the right. Particles may
enter the system at the left and leave at the right. In the bulk all
particles will move with their maximum possible speed, which is either
given by the speed limit $v_{\rm max}$, or it is given by the
distance to the next particle to avoid collisions. There will be no
stochasticity in the bulk and particles always achieve their maximum 
possible speed instantaneously. In the case of periodic boundary
conditions this model is known as the deterministic Fukui-Ishibashi
model \cite{FI96}, for which some exact results are known \cite{F99}. 

With open boundaries, particles will be allowed to enter the system on
the first $v_{\rm max}$ sites and may leave the system from the last
$v_{\rm max}$ sites. The choice of boundary conditions can have a
profound influence on the behaviour of the system. If, for example,
particles were allowed to enter only at the first site, the density
profile for $v_{\rm max}>1$ in the free flow phase would show a strong
sublattice dependence. Moreover, in this case the system would not be
able to reach its maximum possible flow, since particles would have to
wait an extra time step due to the synchronous update before the first
site is unblocked. In the case of $v_{\rm max}=2$, the specific
boundary conditions we will use here are similar to those of the
random sequential model A of \cite{KS99}. If the two sites at the left
boundary are empty, a particle can enter on the second site with
probability $\alpha_2$, and on the first site with probability
$\alpha_1 (1-\alpha_2)$. The sites remain empty with probability
$(1-\alpha_1) (1-\alpha_2)$. If a particle is already present on the
second site, a probability $\alpha_3$ is given for a particle entering
on the first site. At the right boundary, a particle at the last site
will leave the system with probability $\beta_1$. If the last site is
empty, but a particle is present on the penultimate site, it will
leave the system with probability $\beta_2$. In terms of Boolean
variables $\tau_i$ that have the value $1$ for a particle and $0$ for
a hole, the dynamical rule for the bulk can be written as,    
\begin{equation}
\tau_i' = \tau_{i-2} \sigma_{i-1} \sigma_i + \tau_{i-1}
\sigma_i \tau_{i+1} + \tau_i \tau_{i+1}, \label{eq:bulkdyn}
\end{equation}
where the prime denotes time incremented by one and $\sigma =
1-\tau$. At the boundaries additional Boolean variables $\hat{\alpha}
and \hat{\beta}$ are used that have time averages equal to $\alpha$
and $\beta$. At the left boundary the rules are, 
\begin{eqnarray}
\tau_1' &=& \tau_1 \tau_2 + \hat{\alpha}_1  (1-\hat{\alpha}_2)
\sigma_1 \sigma_2 + \hat{\alpha}_3 \sigma_1 \tau_2,\\
\tau_2 ' &=& \tau_1 \sigma_2 \tau_3 + \tau_2 \tau_3 + \hat{\alpha}_2
\sigma_1 \sigma_2,
\end{eqnarray}
and at the right boundary we have,
\begin{equation}
\tau_L' = \tau_{L-2} \sigma_{L-1} \sigma_L + (1-\hat{\beta}_1)
\tau_{L} + (1-\hat{\beta}_2) \tau_{L-1} \sigma_L. \label{eq:rbounddyn}
\end{equation}
The currents are defined by the continuity equation,
\begin{equation}
\tau'_i = \tau_i + j_{i-1} - j_i,
\end{equation}
and are given by,
\begin{eqnarray}
j_0 &=& \hat{\alpha}_3 \sigma_1\tau_2
+ (1-(1-\hat{\alpha}_1)(1-\hat{\alpha}_2)) \sigma_1\sigma_2
\nonumber\\
j_1 &=& \tau_1\sigma_2 +\hat{\alpha}_2\sigma_1\sigma_2\nonumber \\
j_i &=& \tau_i\sigma_{i+1} + \tau_{i-1}\sigma_i\sigma_{i+1} \\
j_L &=& \hat{\beta}_1\tau_L + \hat{\beta}_2 \tau_{L-1}\sigma_L
\nonumber.
\end{eqnarray}
For technical convenience we will put $\langle \hat{\beta}_1 \rangle =
\langle \hat{\beta}_2 \rangle = \beta$, and $\langle \hat{\alpha}_1
\rangle = \langle \hat{\alpha}_2 \rangle = \langle \hat{\alpha}_3 \rangle =
\alpha$ in the rest of this paper. All arguments however hold for the
more general case. The calculation becomes more cumbersome and one has
to discriminate between even and odd sublattices.

\section{The stationary state}
\label{se:state}
In the following discussion, the relative weight of a particular
configuration $\{\tau_1,\ldots,\tau_L\}$ in the stationary state will be
denoted by $P(\tau_1,\ldots,\tau_L)$. Once all relative weights are
determined, the normalisation $Z_L$ can be calculated via,
\begin{equation}
Z_L = \sum_{\{\tau\}} P(\tau_1,\ldots,\tau_L).
\end{equation}
To derive some general conclusions about stationary state, it is
helpful to consider the extreme cases $\alpha=1$ and $\beta=1$ first.
 
\subsection{Free flow}
Because we are considering a deterministic model, spontaneous jams do
not occur. Jams will only build up from obstacles at the right
boundary. Pure free flow configurations are obtained by removing these
obstacles, i.e., $\beta=1$. The dynamical rule at the right boundary
then becomes, 
\begin{equation}
\tau_L' = \tau_{L-2} \sigma_{L-1} \sigma_L.
\end{equation}
Since there will be no jams in the stationary state, its
bulk dynamics is given by,
\begin{equation}
\tau'_i = \tau_{i-2},
\end{equation}
and it follows that the master equation can be written as,
\begin{equation}
P(\tau_1,\ldots,\tau_L) = \sum_{\mu,\mu'} \pf(\tau_1\tau_2\tau_3)
\pf(\tau_2\tau_3\tau_4) P(\tau_3,\ldots,\tau_L,\mu,\mu'),
\end{equation}
where,
\begin{equation}
\renewcommand{\arraystretch}{1.4}
\begin{array}{rclrcl}
\pf(000) &=& (1-\alpha) & \pf(100) &=& \alpha \\
\pf(010) &=& 1 & \pf(001) &=& 1\\
\pf(\tau\tau'\tau'') &=& 0\quad{\rm otherwise}.
\end{array}
\label{eq:pfdef}
\end{equation}
This equation can be solved by the Ansatz,
\begin{equation}
P(\tau_1,\ldots,\tau_L) = R(\tau_{L-1}\tau_L) \prod_{i=1}^{L-2}
\pf(\tau_{i}\tau_{i+1}\tau_{i+2}),
\end{equation}
and we find,
\begin{equation}
R(00) =1,\quad R(10) = R(01) = \alpha.
\end{equation}

\subsection{Jammed flow}
Pure jammed flow configuration are obtained by setting $\alpha=1$. 
From the dynamical rules it follows that in this case configurations
with the sequence $000$ in it do not occur in the stationary
state. This means that the bulk and left boundary dynamics may be
replaced by the simple rule,
\begin{equation}
\tau'_i = \tau_{i+1}.
\end{equation}
The master equation for this case is,
\begin{equation}
P(\tau_1,\ldots,\tau_L) = \sum_\mu \pj(\tau_{L-2}\tau_{L-1}\tau_L)
P(\mu,\tau_1,\ldots,\tau_{L-1}),
\end{equation}
where,
\begin{equation}
\renewcommand{\arraystretch}{1.4}
\begin{array}{rclrcl}
\pj(100) &=& \beta & \pj(010) &=& \beta\\
\pj(001) &=& 1 & \pj(110) &=& \beta\\
\pj(101) &=& 1-\beta & \pj(011) &=& 1-\beta\\
\pj(111) &=& 1-\beta & \pj(\tau\tau'\tau'') &=& 0\quad {\rm otherwise}.
\end{array}
\label{eq:pjdef}
\end{equation}
Again this equation can be solved by a simple Ansatz,
\begin{equation}
P(\tau_1,\ldots,\tau_L) = L(\tau_1\tau_2) \prod_{i=1}^{L-2}
\pj(\tau_i\tau_{i+1}\tau_{i+2}).
\end{equation}
In this case we find,
\begin{equation}
L(00) = \beta^2,\quad L(10) = L(01) = \beta,\quad L(11) =
1-\beta.
\end{equation}

\subsection{The general case}
As mentioned already above, spontaneous jams will not occur since we
are considering a deterministic model. Following a similar line of
reasoning as in \cite{TE98}, this can be deduced from the microscopic
dynamics (\ref{eq:bulkdyn})-(\ref{eq:rbounddyn}).

\begin{enumerate}
\item[i]
The sequence $1100$ can only arise from the same sequence shifted by
one lattice unit, 
\begin{equation}
(\tau_i \tau_{i+1} \sigma_{i+2} \sigma_{i+3})' = \tau_i' \tau_{i+1}
\tau_{i+2} \sigma_{i+3} \sigma_{i+4}.
\end{equation}
Since,
\begin{equation}
(\tau_{L-3} \tau_{L-2} \sigma_{L-1} \sigma_L)''' \sim
(\tau_{L-2}\tau_{L-1}\sigma_L)'' \sim (\tau_{L-1}\tau_L)' = 0,
\end{equation}
it follows that configurations with the sequence $1100$ do not occur
in the stationary state.

\item[ii]
Similarly, a sequence $10100$ can only arise from the same sequence
shifted by one lattice unit or from a sequence with $1100$ in it,
\begin{eqnarray}
(\tau_i\sigma_{i+1} \tau_{i+2} \sigma_{i+3} \sigma_{i+4})' &=&
(\tau_{i-2} \sigma_{i-1} + \tau_{i-1} \sigma_i) (\tau_{i+1}\sigma_{i+2}
\tau_{i+3} \sigma_{i+4} \sigma_{i+5}) \nonumber\\
&& {} + \tau'_i \sigma_{i+1} \tau_{i+2} \tau_{i+3} \sigma_{i+4}
\sigma_{i+5}.
\end{eqnarray}  
Since,
\begin{equation}
(\tau_{L-4}\sigma_{L-3}\tau_{L-2}\sigma_{L-1}\sigma_L)''' \sim
(\tau_{L-2}\sigma_{L-1} \tau_L)' =0,
\end{equation}
it follows with the previous observation that also configurations with
the sequence $10100$ in it do not occur in the stationary state.
\end{enumerate}

It thus follows that each configuration can be divided into three
parts. The first part is a free flow part where there are {\em at
least} two holes between successive particles. This part ends at site
$f$ which denotes the last site of the last $000$ sequence of a
configuration. The dynamics for this part is given by,
\begin{equation} 
\tau'_i = \tau_{i-2},\quad 3 \leq i \leq f.
\end{equation}
The third part starts at site $j$ which denotes the first site of
the first jammed configuration, i.e., a $11$ or a $101$ sequence,
whichever comes first. This part is a jammed flow part where there are
{\em at most} two holes between successive particles. For this part the
dynamics is,
\begin{equation} 
\tau'_i = \tau_{i+1},\quad j \leq i \leq L-1.
\end{equation}
In between these two parts there may be a sequence of $100$'s of
arbitrary length. A general configuration may thus be written as
\begin{equation}
\tau_1\ldots\tau_f(100)^n\tau_j\ldots\tau_L.
\end{equation}
This analysis can be performed in a similar way for models with higher
$v_{\rm max}$. The free flow part will end with $v_{\rm max}+1$ zeros
and the intermediate part will consist of a sequence of blocks, where
each block starts with a $1$ followed by $v_{\rm max}$ zeros. The
jammed flow part starts with any of the local jammed
configurations. These are those sequences where there are less than
$v_{\rm max}$ zeros in between two $1$'s.

The master equation for the stationary state can be written explicitly
in this notation. In the case where the jammed flow starts with a $11$
pair it is given by,
\begin{eqnarray}
&&P(\tau_1\ldots\tau_f(100)^n11\tau_{j+2}\ldots\tau_L) =
\pf(\tau_1\tau_2\tau_3) \pf(\tau_2\tau_3\tau_4)
\pj(\tau_{L-2}\tau_{L-1}\tau_L)
\times \nonumber \\ 
&& \hphantom{P(\tau_1\ldots\tau_f}
 \left[ \vphantom{\sum_{p=1}^n}
P(\tau_3\ldots\tau_f(100)^{n+1}11\tau_{j+2}\ldots\tau_{L-1})
\right. \nonumber\\  
&& \hphantom{P(\tau_1\ldots\tau_f\left[ \vphantom{\sum_{p=1}^n}
\right.} 
{}+ \sum_{p=0}^n
P(\tau_3\ldots\tau_f(100)^{p}001(100)^{n-p}11\tau_{j+2}\ldots\tau_{L-1})
\nonumber\\
&&\hphantom{P(\tau_1\ldots\tau_f\left[ \vphantom{\sum_{p=1}^n}
\right.}
\left. + \sum_{p=0}^n
P(\tau_3\ldots\tau_f(100)^{p}010(100)^{n-p}11\tau_{j+2}\ldots\tau_{L-1})
\right]. \label{eq:ma11}
\end{eqnarray}
A slightly different equation is obtained when the jammed flow starts
with a $101$ sequence,
\begin{eqnarray}
&&P(\tau_1\ldots\tau_f(100)^n101\tau_{j+3}\ldots\tau_L) =
\pf(\tau_1\tau_2\tau_3) \pf(\tau_2\tau_3\tau_4)
\pj(\tau_{L-2}\tau_{L-1}\tau_L)
\times \nonumber \\ 
&& \hphantom{P(\tau_1\ldots\tau_f}
 \left[ \vphantom{\sum_{p=1}^n}
P(\tau_3\ldots\tau_f(100)^{n+1}101\tau_{j+3}\ldots\tau_{L-1})
\right. \nonumber\\  
&& \hphantom{P(\tau_1\ldots\tau_f\left[ \vphantom{\sum_{p=1}^n}
\right.} 
{}+ \sum_{p=0}^n
P(\tau_3\ldots\tau_f(100)^{p}001(100)^{n-p}101\tau_{j+3}\ldots\tau_{L-1})
\nonumber\\
&&\hphantom{P(\tau_1\ldots\tau_f\left[ \vphantom{\sum_{p=1}^n}
\right.}
\left. + \sum_{p=0}^n
P(\tau_3\ldots\tau_f(100)^{p}010(100)^{n-p}101\tau_{j+3}\ldots\tau_{L-1})
\right. \nonumber\\
&&\hphantom{P(\tau_1\ldots\tau_f\left[ \vphantom{\sum_{p=1}^n}
\right.}
\left. + 
P(\tau_3\ldots\tau_f(100)^{n+1}011\tau_{j+3}\ldots\tau_{L-1})
\right]. \label{eq:ma101}
\end{eqnarray}
Similar equations are obtained when $f$ and/or $j$ are close to the
boundary. 

To solve (\ref{eq:ma11}) and (\ref{eq:ma101}) we will employ the
powerful matrix product method \cite{DEHP93}. The relative
probabilities for the stationary are written as,
\begin{equation}
P(\tau_1,\ldots,\tau_L) = \langle \L(\tau_1\tau_2) | \prod_{i=1}^{L-2}
\M(\tau_i \tau_{i+1} \tau_{i+2}) | \R(\tau_{L-1}\tau_L) \rangle,
\label{eq:mp}
\end{equation}
Because of the specific form of each configuration, the following
triangular form for the matrices ${\mathcal M}$ suggests itself,
similar to the $v_{\rm max}=1$ case \cite{dGN99},   
\begin{equation}
\M(\tau\tau'\tau'') = 
\left( \begin{array}{cc}
\pf(\tau\tau'\tau'') F & \S(\tau\tau'\tau'')\\
0 & \pj(\tau\tau'\tau'') J
\end{array}\right), \label{eq:mpa}
\end{equation}
where $\pf(\tau\tau'\tau'')$ and $\pj(\tau\tau'\tau'')$ are defined by
(\ref{eq:pfdef}) and (\ref{eq:pjdef}) respectively. The matrices $F$
and $J$ are yet to be determined. While for $v_{\rm max}=1$
they are just scalars given by $F=\beta$ and $J=\alpha$, they will be
more complicated in the present case. The matrices
$\S(\tau\tau'\tau'')$ will solve the dynamical equations for the
bulk. They  are defined on the interface only and $\S(000) = \S(110) =
\S(101) = \S(011) = \S(111) =0$. A similar decomposition as in
(\ref{eq:mpa}) will be used for the boundary vectors,
\begin{equation}
\langle \L(\tau_1\tau_2)| = \left( \langle \L_{\rm F}(\tau_1\tau_2)|,
\langle \L_{\rm J}(\tau_1\tau_2)| \right),  
\end{equation}
and likewise for $|\R(\tau_{L-1}\tau_L)\rangle$.

To make the following more transparent, we will use the notation
$\S_1=\S(100)$, $\S_2=\S(010)$ and $\S_3=\S(001)$. Upon substitution
one quickly concludes that (\ref{eq:ma11}) and (\ref{eq:ma101}) are
equivalent if, 
\begin{equation}
\S_2 J = \alpha F(\S_2 + (1-\alpha)\S_3). \label{eq:10equiv}
\end{equation}
Let us assume that this relation is satisfied and concentrate on
(\ref{eq:ma11}). Substituting (\ref{eq:mpa}) in (\ref{eq:ma11}) one
finds that,
\begin{eqnarray}
&& \langle \L_{\rm F}(00)| = \langle \L_{\rm F}(10)| =
\langle \L_{\rm F}(01)| = \langle \L_{\rm F}|, \nonumber\\
&& |\R_{\rm J}(00)\rangle = |\R_{\rm J}(10)\rangle = |\R_{\rm J}(01)\rangle
= |\R_{\rm J}(11)\rangle = |\R_{\rm J}\rangle, \label{eq:bulkbc}
\end{eqnarray} 
and that the bulk matrices must satisfy,
\begin{eqnarray}
&& \sum_{p=0}^{n-1}  \alpha^p \beta^{2(n-p-1)} F^{3p+2} \left( \beta^2
\S_3 J^2 + \beta F \S_2 J + F^2 \S_1 \right) J^{3(n-p)-1} + \alpha^n
F^{3n+2} \S_3 J = \nonumber\\ 
&& \hphantom{\sum_{p=0}^{n-1}  \alpha^p \beta} 
\sum_{p=0}^{n}  \alpha^p \beta^{2(n-p)} F^{3p} \left( \beta^2 \S_3
J^2 + \beta F \S_2 J + F^2 \S_1 \right) J^{3(n-p)+1} + \alpha^{n+1}
F^{3(n+1)} \S_3 \nonumber\\
&&\hphantom{\sum_{p=0}^{n-1}  \alpha^p \beta}
{}+ (1-\alpha)^2 (1-\beta) \sum_{p=0}^{n}  \alpha^p \beta^{2(n-p)}
F^{3p+2} \S_3 J^{3(n-p)+1}\nonumber\\
&&\hphantom{\sum_{p=0}^{n-1}  \alpha^p \beta}
{}+ (1-\alpha) (1-\beta) \sum_{p=0}^{n}  \alpha^p \beta^{2(n-p)}
F^{3p+1} \left( \beta \S_3 J + F \S_2 \right) J^{3(n-p)+1}.
\label{eq:bulk}
\end{eqnarray}
%
The requirement that the four sums in (\ref{eq:bulk}) cancel term by
term leads to the following equation,
\begin{eqnarray}
&&F^2 \left( \beta^2 \S_3 J^2 + \beta F \S_2 J + F^2 \S_1 \right) =
\beta^2 \left(\beta^2 \S_3 J^2 + \beta F \S_2 J + F^2 \S_1 \right) J^2
\nonumber\\ 
&& \hphantom{F^2 \S_3 J =}
{}+\beta^2 (1-\alpha)(1-\beta) F \left( (1-\alpha) F
\S_3 + \beta \S_3 J + F \S_2 \right) J^2, \label{eq:bulka}.
\end{eqnarray}
For the the remaining terms in (\ref{eq:bulk}) to cancel, the
following equation must be satisfied,
\begin{eqnarray}
&&F^2 \S_3 J = \alpha F^3 \S_3 + \left( \beta^2 \S_3 J^2 + \beta F
\S_2 J + F^2 \S_1 \right) J \nonumber\\
&& \hphantom{F^2 \S_3 J =}
{} + (1-\alpha)(1-\beta) F \left( (1-\alpha) F
\S_3 + \beta \S_3 J + F \S_2\right) J. \label{eq:bulkb}
\end{eqnarray}
Altogether we get three relations, (\ref{eq:10equiv}),
(\ref{eq:bulka}) and (\ref{eq:bulkb}). These can be rewritten as,    
\begin{eqnarray}
\S_2 J &=& \alpha F (\S_2 + (1-\alpha)\S_3), \nonumber\\
0 &=& \beta (\alpha \beta \S_3 + \S_2) J + F \S_1 \label{eq:bulkrel}\\
\alpha F^2 \S_3 J &=& \alpha \beta^2 \S_3 J^3 + F (
(1-\alpha)(1-\beta) \S_2 + \alpha\beta( 1-\alpha-\beta) \S_3 ) J^2
\nonumber\\
&&{}+ \alpha^2 F^3 \S_3.\nonumber
\end{eqnarray}
Besides these bulk relations, there are boundary relations that follow
from considerations of cases where $f$ and/or $j$ is close to the
boundary. They are not particularly illuminating and are listed in
appendix \ref{se:boundary}. It is important to note that if we
manage to solve these boundary relations such that (\ref{eq:bulkbc})
is also satisfied, a solution of (\ref{eq:bulkrel}) then ensures
stationarity of the matrix  product state (\ref{eq:mp}) for arbitrary
system sizes. This also means that (\ref{eq:bulkbc}) and (\ref{eq:bulkrel})
enable us to extrapolate knowledge of small systems to arbitrary large
ones, which helps us to find a solution of the relations we have
obtained. To find a representation for the matrices $F$ and $J$, we
employ the usual strategy for these type of problems: to consider
explicit solutions for small systems and to try to find relations
between them. Using the Ansatz (\ref{eq:mpa}) for the particular form
of the matrices, we then deduce that the following algebraic relations
hold,  
\begin{eqnarray}
0 &=& F^3 - \beta (1-\alpha-\alpha\beta)F^2 - \alpha\beta^2(2-\alpha)F -
\alpha^2\beta^4 \nonumber\\ 
&=& (F-\beta)(F+\alpha\beta)(F+\alpha\beta^2) - \alpha\beta^2
(1-\alpha) (1-\beta) F. \label{eq:Fcub}
\end{eqnarray}
while $J$ satisfies,
\begin{eqnarray}
0 &=& J^3 - \alpha (2-\alpha)J^2 + \alpha^2 \beta (1-\alpha-\alpha\beta)J + 
\alpha^4\beta^2 \nonumber\\
&=& (J-\alpha)(J-\alpha\beta)(J-\alpha^2\beta) +
\alpha(1-\alpha)(1-\beta) J^2. \label{eq:Jcub}
\end{eqnarray}
Considered as polynomials, each of these equations has three
solutions. These may be thought of as being the eigenvalues of $F$ and
$J$ respectively. One thus finds a three dimensional representation for
$F$ and $J$ for which we have to check that it is compatible with the
other relations. This is indeed the case and two examples of explicit
representations for all objects are given in appendix
\ref{se:repre}. For $\alpha,\beta \neq 0$, $F$ will have non-zero
eigenvalues and  thus is invertible. It is then found that
(\ref{eq:Jcub}) is satisfied by choosing $J=-\alpha^2 \beta^2 F^{-1}$.  

Although the solutions of the cubic equation (\ref{eq:Fcub}) are awkward
expressions in terms of $\alpha$ and $\beta$, they are all real for $0
\leq \alpha,\beta \leq 1$. Let $\lambda_n$ denote the eigenvalues of
$F$ and $\mu_n=-\alpha^2\beta^2/\lambda_n$ the eigenvalues of $J$. They
can be written in the following way, 
\begin{eqnarray}
\lambda_n &=& a + \rho \cos \left( \frac{\phi+2\pi n}{3} \right),\\
\mu_n &=& b + \rho'\sin \left(
\frac{\phi'-(2n+1)\pi}{3}\right).
\end{eqnarray}
with $\lambda_1 < \lambda_2 < 0 < \lambda_3$ and $\mu_3 < 0 < \mu_1 <
\mu_2$ for $0 < \alpha,\beta <1$. Here, $a,b,\rho,\rho',\phi$ and
$\phi'$ are real functions of $\alpha$ and $\beta$ defined in appendix
\ref{se:diagrep}. 

\section{Results}
\label{se:results}
In this section some exact results concerning the phase diagram are
calculated. In the case of the purely free flow and jammed flow
phases, the current and density profiles can be calculated easily. For
general values of the boundary rates the correlation length is
calculated and it is shown that it diverges on a special line in the
phase diagram.

\subsection{Free and jammed flow}
We have seen that the stationary state in both the free flow case
($\beta=1$) and the jammed flow case ($\alpha=1$) is a simple
product. This means that correlations are absent and the density
profile is flat in both cases. The values of the current and the
density are easily calculated and given by,
\begin{eqnarray}
\rho_{\rm F} &=& \frac{\alpha}{1+2\alpha},\quad j_{\rm F} \; =\;
\frac{2\alpha}{1+2\alpha}, \\
\rho_{\rm J} &=& \frac{1-\beta}{1-\beta^3},\quad j_{\rm J} \; =\;
\frac{\beta(1-\beta^2)}{1-\beta^3}.
\end{eqnarray}
It follows that in the free flow phase $j_{\rm F}=2\rho_{\rm F}$. As
expected all particles move with their maximum speed $v_{\rm max} =2$. In
the jammed flow phase, the fundamental diagram is given by $j_{\rm
J}=1-\rho_{\rm J}$. In this case all holes move with their maximum
speed which is equal to $1$. Note that the values of the two 
currents are equal if $2\alpha = \beta(1+\beta)$. We will see in the
next section that this line is the coexistence line in the general
phase diagram.

\subsection{The general case}
To facilitate summing over the stationary state
probabilities, the local free and jammed flow configuration
weights $p_{\rm F}(\tau\tau'\tau'')$ and $p_{\rm F}(\tau\tau'\tau'')$
are collected into matrices $P_{\rm f}$ and $P_{\rm j}$ with
elements,
\begin{eqnarray} 
\left(P_{\rm f,j}\right)_{\tau\tau',\tau'\tau''} &=&
p_{\rm f,j}(\tau\tau'\tau''),\\
&=& 0 \quad {\rm otherwise}.\nonumber
\end{eqnarray}
Likewise, the matrix $S$ and vectors $\langle L|$ and $|R\rangle$ are
defined,
\begin{equation}
S_{\tau\tau',\tau'\tau''} = \S(\tau\tau'\tau''),\quad \langle
L|_{\tau\tau'} = \langle \L(\tau\tau')|,\quad |R\rangle_{\tau\tau'} =
|R(\tau\tau') \rangle.
\end{equation}
The advantage of this notation is that for example the normalisation $Z_L$
can be written compactly as, 
\begin{equation}
Z_L = \sum_{\{\tau\}} P(\tau_1,\ldots,\tau_L) = \langle L| M^{L-2}
|R\rangle, 
\end{equation}
where $M$ is a $24 \times 24$ matrix given by,
\begin{equation}
M = \left(\begin{array}{@{}cc} 
\displaystyle P_{\rm f} \otimes F & \displaystyle S \\
0 & \displaystyle P_{\rm j} \otimes J
\end{array}\right).
\end{equation}
To perform the calculations, it is worth mentioning the following two
intermediate results, 
\begin{eqnarray}
\langle L_{\rm F}| \left(P_{\rm f}\otimes F\right)^n &=&
\alpha\beta^2 (1-\alpha)(1-\beta) (1,1,1,0) \otimes
\left(\frac{\lambda_1^{n+1}}{\lambda_1-\beta},
\frac{\lambda_2^{n+1}}{\lambda_2-\beta},
\frac{\lambda_3^{n+1}}{\lambda_3-\beta}\right),\\
\left( \left(P_{\rm j}\otimes J\right)^{n} |R_{\rm J}\rangle
\right)^t &=& - \frac{\beta}{\alpha \Delta} (1,1,1,1) \otimes \left(
\mu_1^{n+2}, \mu_2^{n+2}, \mu_3^{n+2} \right),
\end{eqnarray}
where the discriminant $\Delta$ is defined by,
\begin{equation}
\Delta = (\lambda_1-\lambda_2)
(\lambda_2-\lambda_3)(\lambda_3-\lambda_1). \label{eq:discr}
\end{equation}
After some laborious manipulations, the normalisation is then found to
be,
\begin{eqnarray}
Z_L &=& \frac{\beta^2(1-\alpha)(1-\beta)}{\alpha
(2\alpha-\beta(1+\beta))\Delta} \left[ \alpha^2 \beta
(1-\alpha)(1+2\alpha) \sum_{i=1}^3 \frac{\lambda_i+\beta^2}{\lambda_i-\beta} 
(\lambda_{i+1}-\lambda_{i+2}) \lambda_i^L\right. \nonumber\\ 
&&\left. {}+ (1-\beta^3) \sum_{i=1}^3 \frac{(\mu_i -
2\alpha)(\mu_i-\alpha^2)}{(\mu_i-\alpha)^2}(\mu_{i+1}-\mu_{i+2})
\mu_i^{L+1} \right], \label{eq:partsum}
\end{eqnarray}
where $\lambda_{3+i} = \lambda_i$ and $\mu_{i+3} = \mu_i$. This
expression is well defined for all $\alpha$ and $\beta$. To see this,
the only non-trivial case we have to consider is
$2\alpha=\beta(1+\beta)$. All other explicit poles in
(\ref{eq:partsum}) are equivalent to $(1-\alpha)(1-\beta)=0$. At
$2\alpha=\beta(1+\beta)$ the solutions $\lambda_i$ simplify
dramatically. In particular $\lambda_1=-\beta^2$ while $\lambda_2$ and
$\lambda_3$ are the roots of a quadratic equation. Furthermore, we
find that $\lambda_3=\mu_2$ and $\lambda_2=\mu_3$. These values imply
that the expression between brackets in (\ref{eq:partsum}) has a zero
that precisely cancels the pole at $2\alpha=\beta(1+\beta)$.

Unfortunately the calculations with the present notations are
still rather intricate and laborious and so far no other explicit
expressions for the general case have been obtained. The phase
behaviour of the model however is similar to that of the case
$v_{\rm max}=1$ \cite{ERS99,dGN99,AS00}. From the explicit form of the
normalisation it is clear that the correlation length will 
be determined by ratio of its largest contributions and we find,
\begin{eqnarray}
\xi_{\rm F}^{-1} &=& \ln \frac{\lambda_3}{\mu_2} \;=\; \ln
\frac{\lambda_1}{-\beta^2}, \qquad {\rm for}\;
2\alpha < \beta(1+\beta), \label{eq:clF}\\ 
\xi_{\rm J}^{-1} &=&  -\xi_{\rm F}^{-1}, \hphantom{{}\;=\; \ln
\frac{\lambda_1}{-\beta^2}} \qquad {\rm for}\;
2\alpha > \beta(1+\beta). \label{eq:clJ}
\end{eqnarray}
There is a low density phase for $2\alpha < \beta(1+\beta)$ and a
high density phase for $2\alpha > \beta(1+\beta)$. In each case, the
bulk density will have the free and jammed flow value
respectively. At the boundaries there will be exponential corrections
of which the correlation lengths are given by (\ref{eq:clF}) and
(\ref{eq:clJ}). The curve $2\alpha=\beta(1+\beta)$ is a coexisting
curve on which the correlation length diverges. The instantaneous
density profile is a shock profile resulting in an average linear
profile.  Across this curve, the average bulk density has a jump of
size $\rho_{\rm J} - \rho_{\rm F}$.

As we have seen above, the locus of the coexisting curve is obtained
by equating the values of the current for the two extreme cases: the
free and jammed flow phases. This seems to be a general feature of
ASEP's and supports the ideas of Kolomeisky {\em et
al.} \cite{KSKS98} for the case of discrete time and parallel
update. In the present model however, the dependence of the correlation
length on $\alpha$ and $\beta$ does not decouple, as is the case for
$v_{\rm max}=1$ \cite{dGN99} and the random sequential ASEP
\cite{SD93a}. It is therefore quite amazing that the locus of the
coexisting curve still can be obtained by a simple mean field analysis.       

\section{Conclusion}
\label{se:concl}
An exact solution for the stationary state of a deterministic traffic
model with $v_{\rm max}=2$ is presented. Apart from the absence of
symmetry due to the lack of a particle-hole duality, the phase
diagram is qualitatively similar to that of the case with $v_{\rm
max}=1$, as expected. This solution might be a first step towards an
exact solution of a realistic traffic model.

The stationary state is presented in a matrix product form, where the
matrices depend on three sites and are 24 dimensional. The matrix
product method has been shown to work extremely well in those cases
where the matrices are either infinite dimensional or of small finite
dimension. Its shortcomings are obvious when the matrices are of large
finite dimension and the eigenvalues become solutions of polynomials
of high degree. As in the present case, it will still be a tedious
technical exercise to derive exact expressions for expectation values
and correlation functions from the exact solution.

Although many relations between matrix elements have been given, no
proper matrix algebra has been derived. It will be interesting to find
this underlying algebra, which at least should be of degree 4 as
suggested by equation (\ref{eq:bulkrel}). This algebra may provide
more convenient ways of deriving expectation values and correlation
functions than the method used in this paper. The appearance of cubic
roots however will remain. 

An obvious and interesting extension of the model will be to include
stochasticity in the bulk hopping rates. This can be done as in the
Fukui-Ishibashi model \cite{FI96}, but more interesting perhaps will
be the aggressive driver Nagel-Schreckenberg model \cite{S98}, which is
closer to real traffic but might be still simple enough to be
analytically tractable. 

As a final remark one should mention that because of the long range
interaction it is unlikely that these models are integrable in the
sense that there would be and underlying Yang-Baxter relation. The
paradigmatic ASEP however is integrable since it is closely related to
the integrable XXZ spin chain. It would be interesting to compare the
matrix product ground states of non-integrable systems with those of
integrable ones \cite{SW97, S00}.

\section*{Acknowledgement}
This work has been supported by the Australian Research Council. I
thank M.~T. Batchelor and V. Mangazeev for valuable discussions,
and am much indebted to A. Schadschneider for his continuous support
and encouragement.  

\appendix

\section{Boundary relations}
\label{se:boundary}
In the cases where $f\leq 3$ extra boundary relations are needed for
(\ref{eq:mp}) to be the stationary state. The following equations can
be deduced from the master equation in those cases,

\begin{itemize}
\item $f=3$.

\begin{eqnarray}
\alpha \langle \L_{\rm F}| F \S_3 J &=& \alpha \beta^2 \langle \L_{\rm
J}(01)| J^3 + (1-\alpha)(1-\beta) \langle \L_{\rm F}| \S_2 J^2
\nonumber\\ 
&&{} + \alpha\beta( 1-\alpha-\beta) \langle \L_{\rm F}| \S_3 J^2 +
\alpha^2 \langle \L_{\rm F}| F^2 \S_3.  
\end{eqnarray}

\item $f=2$. 

\begin{equation}
\alpha\beta (1-\beta) \langle \L_{\rm J}(01)| J = \beta \langle
\L_{\rm J}(10)| J + \langle \L_{\rm F}| \S_1.
\end{equation}
\begin{eqnarray}
\alpha \langle \L_{\rm F}| \S_3 J &=& \alpha \beta^2 \langle \L_{\rm
J}(01)| J^3  + \alpha (1-\alpha)(1-\beta) (\langle \L_{\rm F}| \S_2 +
(1-\alpha)\langle \L_{\rm F}| \S_3) J \nonumber\\ 
&&{} + \alpha\beta(1-\beta) \langle \L_{\rm J}(01)| J^2 + \alpha^2
\langle \L_{\rm F}| F \S_3. 
\end{eqnarray}

\item $f=1$. 

\begin{equation}
\beta^2 \langle \L_{\rm J}(01)| J^2 + \langle \L_{\rm F}| \left( \beta
\S_2 J + F \S_1 \right) = \beta^2 (1-\beta) \left( (1-\alpha) \langle
\L_{\rm J}(01)| + \langle \L_{\rm J}(10)| \right) J^2, 
\end{equation}
\begin{equation}
\langle \L_{\rm J}(01)| J = \alpha \langle \L_{\rm F}| \S_3 +
(1-\beta) \left( (1-\alpha) \langle \L_{\rm J}(01) + \langle \L_{\rm
J}(10)| \right) J. 
\end{equation}

\item $f=0$. 

\begin{equation}
\langle \L_{\rm J}(11)| = \frac{\alpha}{\beta} (1-\beta) \langle
\L_{\rm J}(01)|, \quad \langle \L_{\rm F}(11)| =0,
\end{equation}
\begin{equation}
\beta^2 \langle \L_{\rm J}(11)| J^2 = \beta \langle \L_{\rm J}(10)|J^2 +
\langle \L_{\rm F}| \S_1 J.
\end{equation}
\end{itemize}

Similarly, when $j\geq L-1$ extra relations are needed. These are,
\begin{itemize}

\item $j=L-1$

\begin{equation}
\S_2 |\R_{\rm J}\rangle = \alpha(1-\alpha) |\R_{\rm F}(00)\rangle.
\end{equation}

\item $j=L$

\begin{eqnarray}
F^2 \S_3 |\R_{\rm J}\rangle &=& \alpha F^3|\R_{\rm F}(00)\rangle +
(\beta^2 \S_3 J^2 +\beta F\S_2 J +F^2\S_1 )|\R_{\rm J}\rangle
\nonumber\\
&&{}+ (1-\alpha)(1-\beta) F((1-\alpha)F\S_3 + \beta \S_3 J+F\S_2)
|\R_{\rm J}\rangle.
\end{eqnarray}

\item $j=L+1$

\begin{equation}
(1-(1-\alpha)^2)F^2|\R_{\rm F}(00)\rangle = \beta (\beta \S_3 J +
F\S_2 +(1-\alpha) F\S_3 )|\R_{\rm J}\rangle. 
\end{equation}
\end{itemize}

\section{Representations}
\label{se:repre}
\subsection{The diagonal representation}
\label{se:diagrep}
In this subsection an explicit representation is given in which $F$
and $J$ are diagonal. Let $\lambda_1,\lambda_2$ and $\lambda_3$ denote
the three solutions of the equation,
\begin{equation}
\lambda^3 - 3a \lambda^2 - 3b\lambda - \alpha\beta^2 c = 0,  
\label{eq:cubic}
\end{equation}
where,
\begin{eqnarray}
a &=& (\lambda_1 + \lambda_2 + \lambda_3)/3 = \beta (1-\alpha
-\alpha\beta)/3.  \nonumber\\
b &=& -(\lambda_1 \lambda_2 + \lambda_2 \lambda_3  + \lambda_3
\lambda_1)/(3\beta^2)  = \alpha(2-\alpha)/3. \\
c &=& \lambda_1 \lambda_2 \lambda_3/\beta^2 = \alpha^2
\beta^2. \nonumber 
\end{eqnarray}
Then, $F$ and $J$ are given by,
\begin{equation}
F = \diag \{\lambda_1,\lambda_2,\lambda_3\},\quad J = -\alpha^2
\beta^2 \diag \{\lambda_1^{-1},\lambda_2^{-1},\lambda_3^{-1}\}
\end{equation}
The matrices $\S_i$ are given by,
\begin{eqnarray}
\S_1 &=& \alpha^2\beta^2
\left(\begin{array}{@{}ccc@{}}
0 & -\lambda_1 -\lambda_2 & \lambda_1 + \lambda_3 \\
\lambda_1 + \lambda_2 & 0 & -\lambda_2 - \lambda_3 \\
-\lambda_1 - \lambda_3 & \lambda_2 + \lambda_3 & 0
\end{array}\right), \nonumber\\
\S_2 &=& (1-\alpha) 
\left(\begin{array}{@{}ccc@{}}
0 & -\lambda_1 \lambda_2 & \lambda_1 \lambda_3 \\
\lambda_1 \lambda_2 & 0 & -\lambda_2 \lambda_3 \\
-\lambda_1 \lambda_3 & \lambda_2 \lambda_3 & 0
\end{array}\right),\\
\S_3 &=& 
\left(\begin{array}{@{}ccc@{}}
0 & \lambda_1 \lambda_2 + \alpha\beta^2 & -\lambda_1 \lambda_3  -
\alpha\beta^2\\ 
-\lambda_1 \lambda_2 - \alpha\beta^2 & 0 & \lambda_2 \lambda_3 +
\alpha\beta^2 \\ 
\lambda_1 \lambda_3 + \alpha\beta^2 & -\lambda_2 \lambda_3 -
\alpha\beta^2 & 0 
\end{array}\right). \nonumber
\end{eqnarray}
The corresponding boundary vectors are given by,
\begin{eqnarray}
\langle \L_{\rm F}| &=& \alpha\beta^2 (1-\alpha)(1-\beta)
\left(\frac{\lambda_1}{\lambda_1-\beta}
,\frac{\lambda_2}{\lambda_2-\beta},
\frac{\lambda_3}{\lambda_3-\beta} \right), \nonumber\\
\langle \L_{\rm F}(11)| &=& 0, \nonumber\\
\langle \L_{\rm J}(10)| &=& \left( (\lambda_1 - \beta)
(\lambda_2 - \lambda_3) ((2\alpha-1)\lambda_1 + \alpha^2\beta),
(\lambda_2 - \beta) (\lambda_3 - \lambda_1) ((2\alpha-1)\lambda_2 +
\alpha^2\beta), \right.\nonumber\\
&& \left. (\lambda_3 - \beta^) (\lambda_1 -
\lambda_2)((2\alpha-1)\lambda_3 + \alpha^2\beta) \right) \\
\langle \L_{\rm J}(01)| &=& \alpha\beta^2 (1-\alpha)(1-\beta)
\left(\frac{\lambda_1 (\lambda_2-\lambda_3)}{\lambda_1+\alpha\beta^2},
\frac{\lambda_2 (\lambda_3-\lambda_1)}{\lambda_2+\alpha\beta^2},
\frac{\lambda_3 (\lambda_1-\lambda_2)}{\lambda_3+\alpha\beta^2} \right),  
\nonumber\\
\langle \L_{\rm J}(11)| &=& \frac{\alpha}{\beta} (1-\beta) \langle L_{\rm
J}(01).\nonumber
\end{eqnarray}
and the right boundary vectors are given by,
\begin{eqnarray}
|\R_{\rm J}\rangle &=& - \frac{\alpha^3\beta^5}{\Delta} \left(
\lambda_1^{-2}, \lambda_2^{-2}, \lambda_3^{-2}\right), \nonumber \\
|\R_{\rm F}(00)\rangle &=& - \frac{\beta}{\Delta} \left(
\lambda_1 (\lambda_2-\lambda_3), \lambda_2 (\lambda_3-\lambda_1),
\lambda_3 (\lambda_1-\lambda_2)\right),\\
|\R_{\rm F}(10)\rangle &=& |R_{\rm F}(01)\rangle = |R_{\rm
F}(11)\rangle =0,\nonumber
\end{eqnarray}
where the discriminant $\Delta$ is defined in (\ref{eq:discr}).

The solutions of (\ref{eq:cubic}) are all real for $0< \alpha,\beta <
1$ and can be written in the following form,
\begin{equation}
\lambda_n = a + \rho \cos \left( \frac{\phi+2\pi n}{3} \right),
\end{equation}
where,
\begin{equation}
\rho = 2\sqrt{a^2+\beta^2 b},\qquad
\phi = \arctan \left(\Delta/C \right).
\end{equation}
The discriminant $\Delta$ and $C$ are given by 
\begin{equation}
\Delta = \sqrt{27\rho^6/16-C^2}, \qquad C = 3\sqrt{3}(2a^3
+\beta^2(c+3a b)).
\end{equation}
Similarly, the solutions $\mu_n = -\alpha^2\beta^2/\lambda_n$ of
(\ref{eq:Jcub}) can be written as, 
\begin{equation}
\mu_n = b + \rho'\sin \left( \frac{\phi'-(2n+1)\pi}{3}
\right),
\end{equation}
where,
\begin{eqnarray}
\rho' &=& 2\sqrt{b^2-\alpha^2 a},\qquad
\phi' \; =\; \arctan \left(C'/\Delta \right).\\
C' &=&  3\sqrt{3} \beta^2 (2b^3-\alpha^2(c+3ab))/\alpha^2.
\end{eqnarray}

\subsection{A simple representation}
\label{se:simprep}
Although the eigenvalues of $F$ and $J$, or equivalently the roots of
(\ref{eq:Fcub}) and (\ref{eq:Jcub}), are awkward expressions, an
example of a representation with simple matrix elements is given by, 
\begin{equation}
F= \beta \left(\begin{array}{@{}ccc@{}}
1 & 0 & \alpha \\
1-\alpha & -\alpha\beta & \alpha (1-\alpha) \\
0 & 1-\beta & -\alpha 
\end{array}\right), \quad
J= \alpha \left(\begin{array}{@{}ccc@{}}
0 & \alpha\beta & 0 \\
\beta & 1-\alpha & 1-\beta \\
0 & 1-\alpha & 1 
\end{array}\right). 
\label{eq:matrep}
\end{equation}
The corresponding representations for $\S_1$, $\S_2$ and $\S_3$ are given
by,
\begin{eqnarray}
\S_1 &=& - \frac{\alpha\beta^2}{1-\beta}
\left(\begin{array}{@{}ccc@{}}
\beta(1-\alpha) & 1-\alpha & 1-\beta \\
\alpha\beta (1-\beta) & \alpha(1-\alpha)(1-\beta) & \alpha
(1-\beta)^2\\
\alpha\beta (1-\alpha) & \alpha(1-\alpha)^2 &
\alpha(1-\alpha)(1-\beta) 
\end{array}\right), \nonumber\\
\\
\S_2 &=& \frac{\beta^2}{1-\beta}
\left(\begin{array}{@{}ccc@{}}
1-\alpha & 1-\alpha & 0 \\
0 & \alpha(1-\alpha)(1-\beta) & 0 \\
\alpha(1-\alpha) & \alpha(1-\alpha)^2 & 0 
\end{array}\right),\quad
\S_3 \; =\; \beta \left(\begin{array}{@{}ccc@{}}
0 & 1-\alpha & 1 \\
0 & -\alpha\beta & 0 \\
0 & \alpha(1-\alpha) & \alpha 
\end{array}\right). \nonumber
\end{eqnarray}
The left boundary vectors in this representation are represented by,
\begin{eqnarray}
\langle \L_{\rm F}| &=& (1,1-\alpha,0), \nonumber\\
\langle \L_{\rm J}(10)| &=& \frac{1}{1-\beta} \left( \beta (1-\alpha),
(1-\alpha)(\alpha+\beta -\alpha\beta), \alpha (1-\beta) \right)
\nonumber\\
\\[-\baselineskip]
\langle \L_{\rm J}(00) &=& 0 ,\quad \langle \L_{\rm J}(01)| =
(0,1-\alpha,1) \nonumber\\
\langle \L_{\rm J}(11)| &=& \frac{\alpha}{\beta} (1-\beta) \langle \L_{\rm
J}(01),\nonumber
\end{eqnarray}
and the right boundary vectors are given by,
\begin{eqnarray}
|\R_{\rm J}\rangle &=& (0,1-\beta,1), \nonumber\\
\\[-\baselineskip]
|\R_{\rm F}(10)\rangle &=& |\R_{\rm F}(01)\rangle = |\R_{\rm
F}(11)\rangle =0,\quad |\R_{\rm F}(00)\rangle = \frac{\beta}{\alpha}
(1,0,\alpha).\nonumber
\end{eqnarray}
This representation is convenient for calculations for small
system sizes. For larger system sizes it is more useful to use a
representation in which $F$ and $J$ are diagonal. The price one has to
pay is that the matrix elements will be more complicated because they
are cubic roots.


\begin{thebibliography}{99}
%
\bibitem{L85} T. Liggett, {\em Interacting Particle Systems},
Springer Verlag (1985).
%
\bibitem{L99} T. Liggett, {\em Stochastic Interacting Systems: Contact,
Voter, and Exclusion Processes}, Springer-Verlag (1999).
%
\bibitem{SZ95} B. Schmittmann and R~.K.~P. Zia,
Statistical mechanics of driven diffusive systems, in {\em
Phase Transitions and Critical Phenomena} Vol 17, C. Domb and
J.~L. Lebowitz eds., Academic Press (1995).
%
\bibitem{SW98} M. Schreckenberg and D.~E. Wolf eds.,
{\em Traffic and Granual Flow '97}, Springer (1998).
%
\bibitem{CSS00} D. Chowdhurry, L. Santen and A. Schadschneider,
Statistical Physics of Vehicular Traffic and Some Related Systems,
{\em Phys. Rep.} {\bf 329}, 199 (2000). 
%
\bibitem{DDM92}
B. Derrida, E. Domany and D. Mukamel, An exact solution of a one
dimensional asymmetric exclusion model with open boundaries,
{\em J. Stat. Phys.} {\bf 69}, 667 (1992).
%
\bibitem{DE93a} B. Derrida and M.~R. Evans, Exact correlation
functions in an asymmetric exclusion model with open
boundaries, {\em J. Phys. I} {\bf 3}, 311 (1993).
%
\bibitem{SD93a} G. Sch\"utz and E. Domany, Phase transitions
in an exactly soluble one-dimensional exclusion process,
{\em J. Stat. Phys.} {\bf 72}, 277 (1993).
%
\bibitem{DEHP93}
B. Derrida, M.~R. Evans, V. Hakim and V. Pasquier, Exact solution of a
1D asymmetric exclusion model using a matrix formulation, {\em
J. Phys. A: Math. Gen.} {\bf 26}, 1493 (1993).
%
\bibitem{HN83} V. Hakim and J.~P. Nadal, Exact results for 2D
directed animals on a strip of finite width, {\em J. Phys. A:
Math. Gen.} {\bf 16}, L213 (1983).
%
\bibitem{AKLT87} I. Affleck, T. Kennedy, E. Lieb and H. Tasaki,
Rigorous results on valence-bond ground states in antiferromagnets,
{\em Phys. Rev. Lett.} {\bf 59}, 799 (1987).
%
\bibitem{KSZ91} A. Kl\"umper, A. Schadschneider and J. Zittartz,
Equivalence and solution of anisotropic spin-1 models and generalized
t-J fermion models in one dimension, {\em J. Phys. A: Math. Gen.} {\bf
24}, L955 (1991). 
%
\bibitem{H96} H. Hinrichsen, Matrix product ground states
for exclusion processes with parallel dynamics, J. Phys. A {\bf 29},
3659 (1996).
%
\bibitem{RSS96} N. Rajewsky, A. Schadschneider and M. Schreckenberg,
The asymmetric exclusion model with sequential update, {\em
J. Phys. A: Math. Gen.} {\bf 29}, L305 (1996).
%
\bibitem{HP97} A. Honecker and I. Peschel, Matrix product states for a
one dimensional lattice gas with parallel dynamics, {\em J. Stat. Phys.}
{\bf 88}, 319 (1997). 
%
\bibitem{RSSS98} N. Rajewsky, L. Santen, A. Schadschneider and
M. Schreckenberg, The asymmetric exclusion process: comparison of
update procedures, {\em J. Stat. Phys.} {\bf 92}, 151 (1998).
%
\bibitem{ERS99} M.~R. Evans, N. Rajewsky and E.~R. Speer, Exact
solution of a cellular automaton for traffic, {\em J. Stat. Phys.}
{\bf 95}, 45 (1999). 
%
\bibitem{dGN99}
J. de Gier and B. Nienhuis, Exact stationary state for an asymmetric
exclusion process with fully parallel dynamics, {\em Phys. Rev. E}
{\bf 59}, 4899 (1999).
%
\bibitem{BPV00}
J. Brankov, N. Pesheva and N. Valkov, Exact results for a fully
asymmetric exclusion process with sequential dynamics and open
boundaries, {\em Phys. Rev. E} {\bf 61}, 2300 (2000).
%
\bibitem{DJLS93} B. Derrida, S.~A. Janowsky, J.~L. Lebowitz and
E.~R. Speer, Exact solution of the totally asymmetric exclusion
simple exclusion process: shock profiles, {\em J. Stat. Phys.} {\bf
73}, 813 (1993).
%
\bibitem{EFGM95} M.~R. Evans, D.~P. Foster, C. Godr\`eche and
D. Mukamel, Asymmetric exclusion model with two species: spontaneous
symmetry breaking, {\em J. Stat. Phys.} {\bf 80}, 69 (1995). 
%
\bibitem{M96} K. Mallick, Shocks in the asymmetric exclusion model
with an impurity, {\em J. Phys. A: Math. Gen.} {\bf 29}, 5375 (1996).
%
\bibitem{HS97} H. Hinrichsen and S. Sandow, Deterministic
exclusion process with a stochastic defect, {\em J. Phys. A:
Math. Gen.} {\bf 30}, 2745 (1997).
%
\bibitem{EKKM98} M.~R. Evans, Y. Kafri, H.~M. Koduvely and D. Mukamel,
Phase separation and coarsening in one-dimensional driven diffusive
systems: Local dynamics leading to long-range Hamiltonians, {\em
Phys. Rev. E} {\bf 58}, 2764 (1998).
%
\bibitem{AHR98} P.~F. Arndt, T. Heinzel and V. Rittenberg, Stochastic
models on a ring and quadratic algebras. The three-species diffusion
problem, {\em J. Phys. A: Math. Gen.} {\bf 31}, 833 (1998).
%
\bibitem{ADR98} F.~C. Alcaraz, S. Dasmahapatra and V. Rittenberg,
$N$-species stochastic models with boundaries and quadratic algebras,
{\em J. Phys. A: Math. Gen.} {\bf 31}, 845 (1998).
%
\bibitem{K99} V. Karimipour, A multi species exclusion model and its
relation to traffic flow, {\em Phys. Rev. E} {\bf 59}, 205 (1999) 
%
\bibitem{FJ99} M.~E. Fouladvand and F. Jafarpour, Multi-species asymmetric
exclusion process in ordered sequential update, {\em J. Phys. A:
Math. Gen.} {\bf 32}, 5845 (1999).
%
\bibitem{RSS00} N. Rajewsky, T. Sasamoto and E.~R. Speer, Spatial
particle condensation for an exclusion process on a ring, {\em Physica A}
{\bf 279}, 123 (2000).
%
\bibitem{LPK97} H.~W. Lee, V. Popkov and D. Kim, Two-way traffic flow:
Exactly solvable model of traffic jam, {\em J. Phys. A: Math. Gen.}
{\bf 30}, 8497 (1997).
%
\bibitem{S94} S. Sandow, Partially asymmetric exclusion process with open
boundaries, {\em Phys. Rev. E} {\bf 50}, 2660 (1994).
%
\bibitem{ER96} F.~H.~L. Essler and V. Rittenberg, Representations of
the quadratic algebra and partially asymmetric diffusion with open
boundaries, {\em J. Phys. A: Math. Gen.} {\bf 29}, 3375 (1996).
%
\bibitem{BECE99} R.~A. Blythe, M.~R. Evans, F. Colaiori and
F.~H.~L. Essler, Exact solution of a partially asymmetric exclusion
model using a deformed oscillator algebra, {\em J. Phys. A:
Math. Gen.} {\bf 33}, 2313 (2000). 
%
\bibitem{Sa99} T. Sasamoto, One-dimensional partially
asymmetric simple exclusion process with open boundaries: orthogonal
polynomials approach, {\em J. Phys. A: Math. Gen.} {\bf 32}, 7109
(1999). 
%
\bibitem{DE97} B. Derrida and M.~R. Evans, in {\em Nonequilibrium
Statistical Mechanics in One Dimension}, ed. V. Privman, Cambridge
University Press (1997).
%
\bibitem{D98} B. Derrida, An exactly soluble non-equilibrium
system: The asymmetric simple exclusion process, {\em Phys. Rep.} {\bf
301}, 65-83 (1998). 
%
\bibitem{S00} G. Sch\"utz, Exactly solvable models for many-body
systems far from equilibrium, in {\em Phase Transitions and Critical
Phenomena} Vol 19, eds. C. Domb and J. Lebowitz, Academic Press (2000).
%
\bibitem{KS97} K. Krebs and S. Sandow, Matrix product eigenstates for
one-dimensional stochastic models and quantum spin chains  {\em
J. Phys. A: Math. Gen.} {\bf 30}, 3165 (1997).
%
\bibitem{RS97} N. Rajewsky and M. Schreckenberg, Exact results for one
dimensional stochastic cellular automata for different types of
updates, {\em Physica A} {\bf 245}, 139 (1997).
%
\bibitem{KS99}
K. Klauck and A. Schadschneider, On the ubiquity of matrix-product
states in one-dimensional stochastic processes with boundary
interactions, {\em Physica A} {\bf 271}, 102 (1999).
%
\bibitem{AS00} T. Antal and G.M. Sch\"utz, Asymmetric exclusion
process with nex-nearest-neighbor interaction: Some comments on
traffic flow and a nonequilibrium reentrance transition, {\it
Phys. Rev. E} {\bf 62}, 83.
%
\bibitem{SS98} A. Schadschneider and M. Schreckenberg, Garden of Eden
states in traffic models, {\em J. Phys. A: Math. Gen.} {\bf 31}, L225
(1998).
%
\bibitem{TE98}
L.~G. Tilstra and M.~H. Ernst, Synchronous asymmetric exclusion
process, {\em J. Phys. A: Math. Gen.} {\bf 31},  5033 (1998).
%
\bibitem{FI96} M. Fukui and Y. Ishibashi, Traffic flow in 1D cellular
automaton model including cars moving with high speed,
{\em J. Phys. Soc. Jpn.} {\bf 65}, 1868 (1996). 
%
\bibitem{F99} H. Fuk\'s, Exact results for deterministic cellular
automata traffic models, {\em Phys. Rev. E} {\bf 60}, 197 (1999).
 %
\bibitem{KSKS98} A.~B. Kolomeisky, G.~M. Sch\"utz,
E.~B. Kolomeisky and P.~J. Straley, Phase diagram of one-dimensional
driven lattice gases with open boundaries {\em J. Phys. A: Math. Gen.}
{\bf 31}, 6911 (1998). 
%
\bibitem{S98} A. Schadschneider, Analytical approaches to cellular
automata for traffic flow: Approximations and exact results, in
\cite{SW98}.  
%
\bibitem{SW97} T. Sasamoto and M. Wadati, Stationary state of
integrable systems in matrix product form, {\em J. Phys. Soc. Jpn.}
{\bf 66}, 2618 (1997).
%
\end{thebibliography}
\end{document}